# A sub-Saturn Mass-Radius Desert for Planets with Equilibrium Temperature Less than 600 K

David G. Russell

*Owego Free Academy, 1 Sheldon Guile BLVD, Owego, NY 13827 USA*
*email:* russelld@oacsd.org

**Abstract**

The sample of exoplanets from the NASA Exoplanet Archive with equilibrium temperature ≤ 600 K and with low uncertainty for both mass and radius measurements is found to have a desert in the mass-radius distribution consistent with predictions from the core-accretion scenario. This *sub-Saturn mass-radius desert* is almost completely barren of any planets with both a mass > 20 $M_\oplus$ and a radius in the range 4.0 – 7.5 $R_\oplus$ for the sample of planets with Teq ≤ 600 K. In contrast, the sample of planets with Teq > 630 K includes a large fraction of planets with mass-radius values that fall into the Teq ≤ 600 K *sub-Saturn mass-radius desert*. The difference between the two populations may result from differences in migration history in the core-accretion scenario.

**~1. Introduction**

The number of confirmed exoplanets has now reached almost 5500 in both the NASA Exoplanet Archive (Akeson et al. 2013) and The Extrasolar Planets Encyclopaedia (Schneider et al. 2011). As the catalog of confirmed exoplanets grows, comparisons of planet population characteristics against the predictions of planetary formation models and post-formation evolutionary mechanisms continue to shape our understanding and present new phenomenon requiring explanations. The core accretion model (Pollack et al. 1996) appears capable of explaining many features of the Solar System and the exoplanet population (Raymond et al. 2020). However, the exoplanet radius distribution contains features such as the observed radius gap at 1.8 $R_\oplus$ (Fulton et al. 2017) that suggest the exoplanet population characteristics are shaped not only by formation channels, but also by post-formation evolutionary mechanisms such as photoevaporation and core powered mass loss (Fulton et al. 2017; Ginzburg et al. 2018; Misener and Schlichting 2021).

One prediction of the core-accretion scenario is that there should be a sub-Saturn desert with few planets having a mass in the range 30 - 100 $M_\oplus$ (Ida & Lin 2004; Mordasini et al. 2009). Evidence for this sub-Saturn desert is found in mass-period and radius-period diagrams for orbital periods < 3 days (Szabó & Kiss 2011; Owen & Lai 2018; Szabó & Kalman 2019; Szabó et al. 2023). Bertaux & Ivanova (2022) found the sub-Saturn mass desert is most depleted in the mass range 32 – 64 $M_\oplus$ for periods < 100 days.

Given the small population of exoplanets that currently have low uncertainty simultaneously in both mass and radius values, characterization of planets found in the sub-Saturn desert is difficult. As a result the sub-Saturn desert has been explored in mass-period and radius-period space (Szabó & Kiss 2011; Owen & Lai 2018; Szabo & Kálmán 2019; Szabó et al. 2023) and also by mass distribution (Bertaux & Ivanova 2022), but only recently in mass-radius parameter space (Thorngren et al. 2023). Otegi et al. (2020) compiled a catalog of planets with mass < 120 $M_\oplus$ that have low uncertainty for the mass and radius values. The planets in the sub-Saturn desert from this sample have radii in the range 4 to 14 $R_\oplus$ which suggests significant variation in composition, structure, conditions, formation channels, and post formation evolution.

In this note evidence is presented that the population of exoplanets with equilibrium temperature (Teq) less than 600 K includes a nearly barren *sub-Saturn mass-radius desert* in the mass-radius parameter space for mass > 20 $M_\oplus$ and radius in the range 4.0 – 7.5 $R_\oplus$. The nearly barren state of the *sub-Saturn mass-radius desert* for the sample with Teq ≤ 600 K does not appear to persist for planets with equilibrium temperatures > 630 K.

This note is organized as follows: The sample selection criteria for the sample of exoplanets considered in this note is described in section 2. Section 3 describes the Teq ≤ 600 K and Teq > 630 K samples, discusses the possible selection effects that may affect the sample distribution, and describes characteristics of the sample that indicate the sub-Saturn mass-radius desert is not a selection effect. Section 4 is a discussion of possible explanations for the *sub-Saturn mass-radius desert* in the context of formation models, and section 5 is the conclusion.

## ~2. Sample Selection Criteria

Russell (2023) compiled a catalog of planets from the NASA Exoplanet Archive with Teq ≤ 600 K or stellar flux in Earth units ($S_\oplus$) ≤ 25 within the uncertainty of the Teq or $S_\oplus$ value. To be included in the catalog, an exoplanet was required to have a mass value with uncertainty ≤ 27% and a radius value with uncertainty ≤ 8%. These uncertainty standards are similar to those used by Otegi et al. (2020). The full catalog of planets meeting these selection criteria, as of August 2023 is 103 exoplanets (Russell 2023). All planets from the catalog of Russell (2023) with a mass < 130 $M_\oplus$ are included in the sample for this note.

For comparison with the Teq ≤ 600 K sample (hereafter $T_{<600}$), the NASA Exoplanet Archive was searched for planets with Teq > 630 K or stellar flux > 28 $S_\oplus$ (hereafter $T_{>630}$), and with planetary mass and radius values meeting the same uncertainty standards as the $T_{<600}$ catalog. Planets were included in the mass range 20 – 130 $M_\oplus$ with no restriction on the radius value. Planets were also selected for the $T_{>630}$ sample with a mass < 20 $M_\oplus$ if the radius was > 4.0 $R_\oplus$. Planets with a radius < 4.0 $R_\oplus$ and a mass < 20 $M_\oplus$ do not directly border the *sub-Saturn mass-radius desert* and therefore are not included in the $T_{>630}$ sample for this note.

## ~3. Results

The mass-radius characteristics of the samples meeting the criteria described in section 2 are discussed below. As the population of exoplanets meeting the strict selection criteria for mass and radius uncertainty continues to grow it will be important to revisit these characteristics. The results presented in this note are considered preliminary based upon the comparatively small sample of currently confirmed exoplanets that meet the selection criteria described in section 2.

### ~3.1 Mass-radius distribution for planets with Teq ≤ 600 K

Figure 1 presents the mass-radius plot for all planets from the $T_{<600}$ catalog (Russell 2023) with a mass < 130 $M_\oplus$. The black box in Figure 1 outlines an almost completely barren *sub-Saturn mass-radius desert* (hereafter *sSMRd*) which is the mass-radius space encompassing both the mass range 20 – 95 $M_\oplus$ and the radius range 4.0 – 7.5 $R_\oplus$. The single planet within the $T_{<600}$ *sSMRd* is the gas-rich super-Neptune TOI-3884 b ($M_\oplus$ = 32.59, $R_\oplus$ = 6.43 – Libby-Roberts et al. 2023).

Several characteristics of the mass-radius distribution in Figure 1, can be understood in the context of the core accretion scenario for planet formation. The low mass end of the *sSMRd* is 20 $M_\oplus$, which corresponds with the core accretion pebble isolation mass (Lambrechts et al. 2014) and the critical mass for runaway gas accretion (Pollack et al. 1996; Ida & Lin 2004, Raymond et al. 2020). For the $T_{<600}$ sample, the population of Neptune composition planets ($f_{H-He}$ 1 – 49 % by mass) almost truncates at 16 $M_\oplus$ (Figure 1). This feature of the $T_{<600}$ sample is consistent with the core accretion prediction that once a critical mass of ~20 $M_\oplus$ is reached, the planet will begin a stage of runaway gas accretion resulting in a massive Gas Giant composition planet with $f_{H-He}$ > 50.0 % by mass and radius > 7.5 $R_\oplus$ using the models of Fortney et al. (2007).

There are six planets in the $T_{<600}$ sample with a mass < 20 $M_\oplus$ but radii in the range of the *sSMRd*. These six planets outline the low mass edge of the *sSMRd*. In the context of the core accretion scenario these gas-rich Neptunes ($f_{H-He}$ 20 – 49 % by mass) did not reach the critical mass for runaway accretion before the gas disk dissipated.

Figure 1 also has a population of $T_{<600}$ planets with radii from 7.8 – 10.0 $R_\oplus$ that includes 14 planets ranging from 14.7 – 127 $M_\oplus$. From the models of Fortney et al. (2007), these planets are Saturn composition Gas Giants ($f_{H-He}$ 50 – 80 % by mass) and outline the upper radius edge of the *sSMRd*. The Saturn composition population in Figure 1, along with the nearly barren state of the *sSMRd*, appears to be consistent with the core accretion prediction that a planet that reaches the critical core mass of ~20 $M_\oplus$ before the gas disk dissipates has reached the crossover gas mass fraction (~50% H-He by mass) and will finish formation as a Gas Giant composition planet with a mass > 20 $M_\oplus$ and a radius > 7.5 $R_\oplus$.

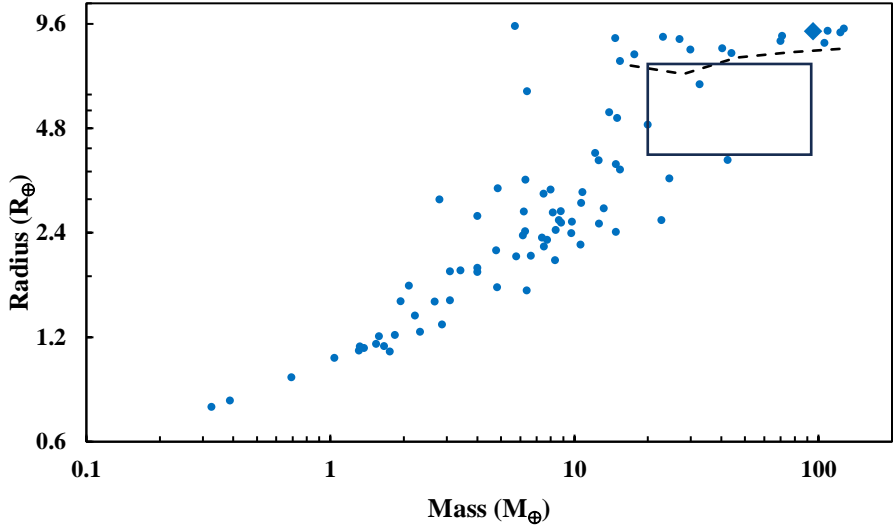

**Fig. 1.** Mass-Radius plot for all exoplanets with Teq ≤ 600 K and mass ≤ 130 M$_⊕$ (blue dots) from the catalog of Russell (2023). Saturn is the blue diamond. The black box outlines the *sub-Saturn mass-radius desert* which encompasses the mass range 20 – 95 M$_⊕$ and the radius range 4.0 -7.5 R$_⊕$. The black dashed line represents an H-He mass fraction of 50% for an equilibrium temperature of 875 K and age of 4.5 Gyr following Fortney et al. (2007).

It should also be noted that there are no Jupiter composition ($f_{H-He}$ 80 – 100 % by mass) Gas Giants in the T$_{<600}$ sample with a mass < 130 M$_⊕$ using the models of Fortney et al. (2007). This is indicated from the lack of planets in the sample with radius > 10.0 R$_⊕$ and mass < 130 M$_⊕$ (Russell 2023). The reason that the Gas Giants with mass < 130 M$_⊕$ and T$_{<600}$ have an upper limit to the $f_{H-He}$ of approximately 80% is an interesting question. This characteristic of the T$_{<600}$ sample could be related to a giant planet formation model with three distinct accretion stages in which H-He runaway accretion actually begins at approximately 100 M$_⊕$ (Helled 2023).

Finally, the sample also contains three planets with mass in the range 20 - 45 M$_⊕$ but radii less than 4.0 R$_⊕$. These Neptune composition planets help to frame a portion of the small radius edge of the *sSMRd*. These planets exceed the critical mass for runaway accretion and may have formed in a gas poor disk environment (e.g. Lee & Chiang 2016).

### ~3.2 Mass-radius distribution for planets with Teq > 630 K

Figure 2 presents the mass-radius plot for the planets from Figure 1 (blue dots) and planets with T$_{>630}$ and radius < 10 R$_⊕$ meeting the selection criteria described in section 2 (red dots). There are 70 planets with T$_{>630}$ plotted in Figure 2. An additional 40 Gas Giants were identified in the NASA Exoplanet Archive with radius > 10 R$_⊕$ and mass < 130 M$_⊕$ which meet the selection criteria for mass and radius uncertainty but are not plotted in Figure 2. Table 1 summarizes the mass-radius distribution for the planets in this sample.

There are 29 planets with T$_{>630}$ that fall into the T$_{<600}$ *sub-Saturn mass-radius desert*. Twenty five of these planets have a mass in the range 20 – 45 M$_⊕$ with the remaining four falling in the mass range 45 – 70 M$_⊕$. There is evidence for a radius break in the T$_{>630}$ *sSMRd* population at a mass of 31 M$_⊕$ as 12/15 planets in the *sSMRd* with M$_⊕$ < 31 have R$_⊕$ < 5.5 whereas 13/14 planets with M$_⊕$ > 31 have R$_⊕$ > 5.5.

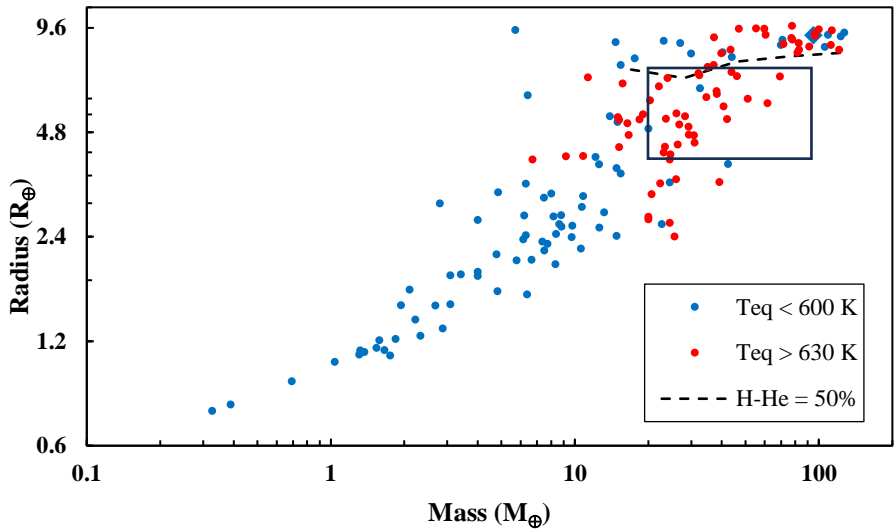

**Fig. 2.** Mass-Radius plot for all exoplanets with Teq ≤ 600 K and mass ≤ 130 $M_\oplus$ (blue dots) from the catalog of Russell (2023). Saturn is the blue diamond. The black box outlines the *sub-Saturn mass-radius desert* which encompasses the mass range 20 – 95 $M_\oplus$ and the radius range 4.0 -7.5 $R_\oplus$. Red dots are planets from the NASA Exoplanet Archive with Teq > 630 K selected as described in section 2. The black dashed line represents an H-He mass fraction of 50% for an equilibrium temperature of 875 K and age of 4.5 Gyr following Fortney et al. (2007).

While the sample of exoplanets meeting the strict mass-radius uncertainty standards is small, the sample indicates that that the nearly barren state of the sSMRd is specific to planets with Teq ≤ 600 K (Table 1). For the sample of planets in the mass range 20 – 130 $M_\oplus$, 29.6% of the $T_{>630}$ planets are in the *sSMRd* whereas only 6.3% of the $T_{<600}$ planets are in the *sSMRd*. If considering only planets with space-based transit data 39.3% of $T_{>630}$ planets are in the *sSMRd* whereas only 6.7% of the $T_{<600}$ planets are in the *sSMRd*.

This can be further characterized in terms of orbital periods. For the $T_{>630}$ *sSMRd* population, 25/29 planets have an orbital period less than 12 days and all have an orbital period less than 25 days. The single $T_{<600}$ planet (TOI-3884 b) in the *sSMRd* also has an orbital period < 12 days. The mean orbital period of the 30 planets in the sSMRd is 6.68 days. In contrast to the sSMRd planets, the eighteen planets with radius in the range 4 – 7.5 $R_\oplus$ but with $M_\oplus$ ≤ 20, have a mean orbital period of 32.18 days with mean orbital periods of 17.93 days for the twelve $T_{>630}$ planets and 60.66 days for the six $T_{<600}$ planets.

3.3 Possible impact of observational selection effects

Given the small number of exoplanets meeting the strict mass and radius uncertainty standards for the sample, selection effects could create false features in the mass-radius distribution and should be kept in mind when considering this sample. Specifically, transit detections are affected by the geometry of the planetary orbit, limitations of ground-based vs. space-based transit surveys, and differences in survey design for the space-based missions such as the Kepler, K2, and TESS surveys.

**Table 1**

| Sample radius-mass ranges | | $N_{tot}$ $T_{eq} < 600$ | $N_{tot}$ $T_{eq} > 630$ | $N_{s\text{-}b}$ $T_{eq} < 600$ | $N_{s\text{-}b}$ $T_{eq} > 630$ |
|---|---|---|---|---|---|
| $R_\oplus$ 4 – 7.5 | $M_\oplus < 20$ | 6 | 12 | 6 | 12 |
| $R_\oplus < 4$ | $M_\oplus$ 20 - 130 | 3 | 8 | 3 | 7 |
| $R_\oplus$ 4 – 7.5 | $M_\oplus$ 20 – 130 | 1 | 29 | 1 | 22 |
| $R_\oplus$ 7.5 – 10 | $M_\oplus$ 20 – 130 | 14* | 21 | 10 | 14 |
| $R_\oplus > 10$ | $M_\oplus$ 20 – 130 | 0 | 40 | 0 | 13 |

Sample population distribution for the planet sub-samples discussed in this note. $N_{tot}$ is the full sample that includes all planets meeting the mass and radius uncertainty standards. $N_{s\text{-}b}$ is the subsample of planets from $N_{tot}$ that have radii determined from space-based transit data. *Total includes Saturn.

The exoplanet sample meeting the selection criteria in this note indicates that the sSMRd mass-radius space identified for the $T_{<600}$ sample is occupied by a large number of short orbital period (period < 12 days) planets with $T_{>630}$. Planets with a short orbital period are more likely to have an observable transit configuration (e.g. Winn 2010; Youdin 2011) and therefore represent a higher fraction of the population of detected exoplanets. In this respect, it is not unexpected to find that a given mass-radius range, such as the *sSMRD*, has more planets with higher equilibrium temperature than with lower equilibrium temperature.

Planets with larger radii will also have a higher detection rate (Youdin 2011). When this is combined with the challenges the Earth's atmosphere presents for ground based transit studies in detecting smaller radius planets (Winn 2010), it is also not unexpected that planets with radius > 10 $R_\oplus$ represent the largest $T_{>630}$ subsample (Table 1). The impact of ground-based transit observations on the sample distribution is illustrated in Table 1 as the number of gas giants with radius > 10 $R_\oplus$ is reduced from 40 to only 13 planets when the ground-based transit sample is removed.

The Kepler, K2, and TESS space-based transit surveys each have different detection limitations which have the potential to severely complicate interpretation of the population distribution in any sample that draws from all three surveys. The original Kepler mission observed a single star field continuously and therefore was capable of detecting planets with long orbital periods (> 100 days) but only within the single field observed (Borucki et al. 2010). The follow-up Kepler K2 mission (Howell et al. 2014) observed multiple target fields near the ecliptic continuously for approximately 75 days, which sets the upper limit for orbital periods that can be reliably confirmed by the K2 mission. Finally, the TESS mission (Ricker et al. 2014; Huang et al. 2018) is an all sky mission that observes each sky sector for two full orbits which sets an upper limit of 27.4 days for the orbital period of detected exoplanets over most of the celestial sphere. However, the TESS mission does have smaller areas of coverage surrounding the ecliptic poles that are observed for approximately 80 days and more than 300 days allowing for longer orbital period planets to be identified in these sectors. Given the variation in the Kepler, K2, and TESS missions and the higher detection rate for planets with shorter orbital periods, the $T_{<600}$ *sSMRd* identified in this note could be an artifact from the lower detection rate for exoplanets with orbital periods greater than 25 days.

Uncertainty in mass and radius values can also present challenges to interpreting exoplanet population statistics. This concern is mitigated to some extent with the sample for this note by restricting the sample to exoplanets with radius uncertainty ≤ 8% and mass uncertainty ≤ 27%. However, even planets with low reported uncertainty in mass and radius can still have significant changes to mass or radius values with follow-up studies. As an example, the single $T_{<600}$ planet currently residing in the mass-radius space of the *sSMRd* is TOI-3884 b, which has mass and radius values of 32.59 +7.31/-7.38 $M_\oplus$ and 6.43 +/- 0.20 $R_\oplus$ respectively (Libby-Roberts et al. 2023). In the initial discovery paper, Almenara et al. (2022) found TOI-3884 b has a mass of 16.5 +3.5/-1.8 $M_\oplus$ and a radius of 6.00 +/-0.18 $R_\oplus$, values that also meet the uncertainty standards of this sample. The updated mass and radius values for TOI-3884 b (Libby-Roberts et al. 2023) are 4.6x and 2.4x larger respectively than the uncertainty in the mass and radius values reported by Almenara et al. (2022).

It is also important to note that the initial mass and radius values for TOI-3884 b (Almenara et al. 2022) would place the planet below the low mass edge of the *sSMRd* and leave the desert completely barren for planets with $T_{<600}$. Kepler 10 c provides another example of a planet with a significant change in estimated mass value with updated data. Dumusque et al. (2014) determined a mass of 17.2 +/- 1.9 $M_\oplus$ for Kepler 10 c but more recently Bonomo et al. (2023)

found a mass of 11.4 +/-1.3 M⊕. TOI-3884 b and Kepler 10 c are examples that illustrate the potential for improved mass and radius values to significantly shift the position of any given planet in mass-radius space.

3.4. The $T_{<600}$ *sSMRd*: Is it real or an artifact from selection effects?

The important question is whether the $T_{<600}$ *sSMRd* is a real phenomenon in the planetary mass-radius distribution or, instead, an artifact from selection effects and data errors such as those outlined in section 3.3. While it is important to keep the selection effects discussed in section 3.3 in mind, the sample of exoplanets plotted in Figures 1 and 2 and summarized in Table 1 does have characteristics that indicate the $T_{<600}$ *sSMRd* is likely a real desert in the exoplanet population. If real, the $T_{<600}$ *sSMRd* could have important implications for understanding the core-accretion scenario for planet formation (section 4).

An important population in this sample is the planets with radii in the range of the $T_{<600}$ *sSMRd*, 4 – 7.5 M⊕, but with a mass less than 20 M⊕. There are 18 planets in this group with twelve in the $T_{>630}$ group and six in the $T_{<600}$ group (Table 1). This group will be identified as "gas-rich Neptunes" for the discussion that follows. All eighteen gas-rich Neptunes have radius values from space-based transit surveys and therefore only exoplanets with space-based transit data will be considered for the remainder of this discussion.

The median orbital periods for the $T_{>630}$ and $T_{<600}$ gas-rich Neptunes are 13.8 (range 6.04 – 52.09) and 45.0 (range 4.67 – 191.23) days respectively. In contrast, the median orbital period for the twenty two $T_{>630}$ planets in the *sSMRd* is only 4.97 (range 0.79 – 24.28) days and the single $T_{<600}$ planet in the *sSMRd* has an orbital period of 4.54 days. Given that smaller mass planets are easier to detect at shorter orbital periods, the more massive planets of the *sSMRd* could be expected to have a larger median orbital period than the gas-rich Neptunes if selection effects are significantly shaping the mass-radius distribution of the two populations. The significantly shorter median orbital period for the planets in the *sSMRd* as compared to the lower mass gas-rich Neptunes suggests that formation mechanisms or post formation evolutionary processes are contributing substantially to the observed mass-radius distribution.

It is also important to note that the $T_{>630}/T_{<600}$ population ratio of the *sSMRd* is significantly different from the other sub-populations in this sample. From the space-based transit sample in Table 1 it can be seen that the gas-rich Neptunes, planets with mass > 20 M⊕ and radius < 4 M⊕, and gas giants with mass in the mass range 20 - 130 M⊕ have $T_{>630}/T_{<600}$ ratios of 2.0, 2.3, and 2.7 respectively. In contrast, the $T_{>630}/T_{<600}$ ratio for the *sSMRd* is 22.0.

As a further check, the Kepler, K2, and TESS catalogs in the NASA Exoplanet Archive were searched for all planets that have radius values with uncertainty ≤ +/- 8% in the radius range 4 – 7.5 R⊕, but with no mass value or mass values that do not meet the uncertainty standard of this sample. This search identified 37 $T_{>630}$ and 17 $T_{<600}$ planets respectively, resulting in a $T_{>630}/T_{<600}$ ratio of 2.2. All sub-samples, except the *sSMRd*, have a $T_{>630}/T_{<600}$ ratio in the narrow range 2.0 – 2.7. With a $T_{>630}/T_{<600}$ ratio of 22.0, the *sSMRd* sub-sample is an outlier and hints at a real phenomenon underlying the difference between the $T_{>630}$ and $T_{<600}$ samples.

The gas-rich Neptune TOI-5678 b (Teq = 513 K) has mass and radius values of 20 +/-4 M⊕ and 4.91 +/- 0.08 R⊕ respectively (Ulmer-Moll et al. 2023). The mass of this planet is exactly on the boundary between the gas-rich Neptune and *sSMRd* populations. With an orbital period of 47.73 days this planet is a better fit with the characteristics of the gas-rich Neptune planet population and has therefore been included in the gas-rich Neptune sample. If TOI-5678 b was shifted to the $T_{<600}$ *sSMRd* population the $T_{>630}/T_{<600}$ ratio for the *sSMRd* would be reduced to 11.0, which is still significantly higher than the $T_{>630}/T_{<600}$ ratio for the other sub-samples.

The significantly larger $T_{>630}/T_{<600}$ ratio for the *sSMRd* indicates either (1) a significant bias or selection effect in the data that specifically is affecting the detection of $T_{<600}$ planets in the mass-radius space of the *sSMRd*, or (2) the presence of a real phenomenon in the exoplanet mass-radius distribution that may result from formation mechanisms or post formation evolution. The gas-rich Neptune population discussed above would seem to support formation or post formation evolutionary mechanisms over selection effects in the data as the reason for the $T_{<600}$ *sSMRd*. Based upon radius, $T_{<600}$ planets in the *sSMRd* should be as easy to detect as the $T_{<600}$ gas-rich Neptunes. Given that the $T_{>630}/T_{<600}$ ratio for the gas-rich Neptunes is 2.0, it is difficult to identify a selection effect in the same radius range that would result in a large deficit in the detection of $T_{<600}$ planets with mass exceeding 20 M⊕ but not for planets with mass less than 20 M⊕.

The most plausible selection effect that could explain the lack of $T_{<600}$ planets in the *sSMRd*, while at the same time allowing for a $T_{>630}/T_{<600}$ ratio of 2 for the gas-rich Neptunes, would be a failure to detect $T_{<600}$ planets in the *sSMRd* because the real population is found at much larger orbital periods than the current surveys can efficiently detect. However, as already noted, the $T_{>630}$ planets in the *sSMRd* have shorter orbital periods than the $T_{>630}$ gas-rich Neptunes even though the greater masses of planets in the *sSMRd* would suggest they should be detectable at larger orbital periods. Where are the massive ($> 20$ $M_⊕$) sub-Saturn radius planets with $T_{<600}$ and orbital periods in the range of the smaller mass $T_{<600}$ gas-rich Neptunes? Even if the population is undetected due to the real population having much longer orbital periods than can be detected from current space-based transit surveys, that difference alone would imply an important difference in the orbital characteristics of the real $T_{<600}$ population of the *sSMRd* compared to the $T_{<600}$ gas-rich Neptune population.

**~4. Discussion**

Restricting the exoplanet sample to planets with low uncertainty in both mass and radius results in significantly smaller sample sizes than available for mass-period and radius-period analyses of the sub-Saturn desert predicted by the core-accretion scenario. However, the sample of planets with low mass and radius uncertainty allows for characterization of the exoplanet distribution in mass-radius space and for characterization of planetary composition classes (e.g. Otegi et al. 2020; Russell 2023). In this note an almost barren *sub-Saturn mass-radius desert* has been identified for planets with $T_{<600}$. The desert is defined by the mass-radius parameter space $20 - 95$ $M_⊕$ and $4.0 - 7.5$ $R_⊕$. Planets with $T_{>630}$ occupy this desert at a substantially higher rate. Most of these planets have orbital periods less than 8 days and all have orbital periods less than 25 days. While currently almost barren, additional planets with $T_{<600}$ in the mass-radius space of the *sSMRd* may be identified with new discoveries and improved data (e.g. TOI-3884 b). The sample presented in this note should be considered preliminary.

Russell (2023) noted several ways that the full $T_{<600}$ exoplanet catalog sample appears consistent with predictions of the core accretion scenario. Specifically, the full catalog sample indicates that: (1) the population of planets that have mass-radius values consistent with a Terrestrial composition lacking an H-He envelope truncates at 3.0 $M_⊕$ which is below the critical core mass for accreting a substantial H-He envelope (Ida & Lin 2004), (2) the critical core mass for accreting a substantial gas envelope appears to fall in the mass range $4.75 - 6.75$ $M_⊕$ consistent with Ida & Lin (2004), (3) the Neptune composition planet population nearly truncates at 18 $M_⊕$ which is consistent with both the critical mass for runaway accretion (Pollack et al. 1996; Ida & Lin 2004; Raymond et al. 2020) and the pebble isolation mass (Lambrechts et al. 2014), (4) There exists a *sSMRd* for masses $< 100$ $M_⊕$ (Ida & Lin 2004; Mordasini et al. 2009).

The $T_{>630}$ sample also has numerous Gas Giant composition planets with radii exceeding 10 $R_⊕$ whereas the $T_{<600}$ sample has none. The large radii of many of these Gas Giants might be explained by a combination of high stellar irradiation and tidal inflation effects (Fortney et al. 2007; Millholland et al. 2020). The orbital periods of the Gas Giants in the two samples support this possibility. For the $T_{>630}$ sample all of the Gas Giants with radii $> 10$ $R_⊕$ have orbital periods less than 12 days and the Gas Giants with radii between 7.9 and 10 $R_⊕$ all have orbital periods less than 20 days. For the $T_{<600}$ sample, only 6 of the 32 Gas Giants have orbital periods less than 28 days. The shorter orbital periods for the $T_{>630}$ sample will lead to higher post-formation inflationary effects from both stellar irradiation and tidal inflation (Fortney et al. 2007; Lopez & Fortney 2013; Millholland et al. 2020).

How can the $T_{>630}$ planets located in the *sSMRd* desert be explained? The differences between the $T_{<600}$ and $T_{>630}$ populations could be an indication that the $T_{<600}$ sample represents the planet population core-accretion produces when planets experience limited post-formation evolutionary effects from high stellar irradiation and tidal action. For example, whereas all but one of the Terrestrial planets from the $T_{<600}$ sample of Russell (2023) has a mass $< 3.0$ $M_⊕$, there are numerous $T_{>630}$ Terrestrial composition planets with much higher masses ($3.5 - 10$ $M_⊕$). These higher mass and high equilibrium temperature Terrestrial composition planets have been interpreted as sub-Neptune cores stripped of their H-He envelopes (e.g. Dai et al. 2019) and therefore are not analogs for the formation pathways of the Solar System Terrestrial planets. In contrast, the $T_{<600}$ Terrestrial planets could be planets that never accreted a substantial H-He envelope and are therefore better analogs for the Solar System Terrestrial planets.

In a similar way, the lack of $T_{<600}$ planets in the sub-Saturn mass-radius desert could indicate that core accretion rarely forms planets that finish with larger orbital periods (>20 days) in this mass-radius space. In this general scenario, the $T_{>630}$ planets found in the *sSMRd* either finished formation occupying the desert, or evolved into the desert from post-formation evolutionary effects. Mechanisms that can account for the $T_{>630}$ planet population in the *sSMRd* should be mechanisms that work more efficiently at higher equilibrium temperature or with migration to shorter orbital periods in order to be consistent with the rarity of such planets with T$_{<600}$.

It is important to note that while the *sSMRd* was initially identified from the sample of planets with T$_{<600}$ (Russell 2023), low equilibrium temperature by itself is unlikely to provide the full mechanism that explains the lack of T$_{<600}$ planets in the desert. This point is underscored by the fact that TOI-3884 b, the single T$_{<600}$ planet in the desert, has an equilibrium temperature of 441 K and an orbital period of 4.54 days (Libby-Roberts et al. 2023). It therefore seems likely that processes related to migration into short orbital periods play a significant role in the formation, or post-formation evolution, of planets that occupy the *sSMRd*. This is also consistent with the gas-rich Neptune population in this sample having longer mean and median orbital periods than the more massive planets with the same radius range found in the *sSMRd*.

Several possible pathways and mechanisms for planets with T$_{>630}$ either forming or evolving into the *sSMRd* are noted below:

~Planets originally formed as Gas Giants that subsequently lost a significant fraction of their H-He envelope through processes such as photoevaporation, radius inflation, Roche lobe overflow, or other mechanisms of hydrodynamical escape (e.g. Lopez & Fortney 2013; Armstrong et al. 2020; Kubyshkina & Fossati 2022; Thorngren et al. 2023).

~Neptune composition planets that have radii inflated from high stellar flux and tidal inflation (e.g. Lopez & Fortney 2014; Millholland et al. 2020).

~Planets that formed in a gas poor environment (e.g. Lee & Chiang 2016) or were in the early stages of runaway gas accretion as the gas disk dissipated and therefore were unable to accrete a massive gas envelope.

~Planets that experienced a significant metallicity increase from planetesimal accretion during migration (e.g. Shibata et al. 2020, 2022).

~Planets that experienced giant impacts during formation (e.g. Ginzburg & Chiang 2020; Ogihara et al. 2021).

~Variations in proto-planetary disk structure, metallicity, pebble and planetesimal supply, and planet migration history (see review of formation mechanisms by Raymond & Morbidelli 2022).

One intriguing possibility is that the T$_{>630}$ planets occupying the *sSMRd* finished formation within the mass-radius space of the desert due to suppression of gas accretion in a hybrid pebble-planetesimal scenario (Kessler & Alibert 2023). In their models, Kessler & Alibert (2023) found that, for most hybrid pebble-planetesimal disk scenarios, planets will migrate inward to approximately 0.05 AU and finish with masses in the range 23 – 70 M$_\oplus$ which is an excellent match for the mass range of the T$_{>630}$ planets occupying the *sSMRd* (Figure 2). This model may be consistent with the near absence of T$_{<600}$ planets in the *sSMRd* and the shorter orbital periods for the T$_{>630}$ planets found in the sSMRd as compared to the longer orbital periods of both the T$_{<600}$ and T$_{>630}$ gas-rich Neptunes.

It is noteworthy that when the T$_{<600}$ and T$_{>630}$ samples are combined, and despite their smaller masses, only 3 of the 18 gas-rich Neptunes have an orbital period less than 8 days whereas 24 of the 30 planets within the *sSMRd* have an orbital period less than 8 days. This difference in orbital periods appears to support the role of migration and the higher metallicity, higher temperature environment close to the star as an important part of the pathways that helped form the T$_{>630}$ planets that occupy the *sSMRd*. It also highlights the need to consider the role that variations in proto-planetary disk composition and structure may have in producing these planets (e.g. Kessler & Alibert 2023).

~5. Conclusion

The preliminary conclusion indicated from the sample of exoplanets currently available is that the mass-radius distribution for planets with equilibrium temperature less than 600 K includes a nearly barren *sub-Saturn mass-radius desert* for the mass range 20 – 95 M$_\oplus$ and the radius range 4.0 – 7.5 R$_\oplus$. This desert is consistent with predictions of the core accretion scenario in the following aspects: (1) the lower mass limit for this desert is consistent with the predicted critical mass for runaway accretion (Pollack et al. 1996; Ida & Lin 2004, Raymond et al. 2020) and the pebble isolation mass (Lambrechts et al. 2014), (2) the mass range of the desert matches the prediction that planets with a mass in the range 30 – 100 $M_\oplus$ should be rare (Ida & Lin 2004; Mordasini et al. 2009), and (3) planets that do reach a mass > 20 $M_\oplus$ will experience runaway accretion and finish with a Gas Giant composition ($f_{H\text{-}He}$ > 50 % by mass – Pollack et al. 1996; Ida & Lin 2004).

The sample of exoplanets with $T_{>630}$ includes numerous planets within the *sSMRd*. The difference between the $T_{<600}$ and $T_{>630}$ samples suggest two broad aspects: (1) differences in formation conditions and migration history, and (2) differences in post-formation evolutionary influences. One possibility is that the planets with $T_{<600}$ form a nearly barren desert because core accretion rarely forms planets within the mass-radius parameter space of the *sSMRd* at larger orbital distances (p > 20 - 25 days). Planets either stall formation below the critical mass for runaway accretion (~20 $M_\oplus$), or upon reaching the critical mass successfully accrete enough H-He to achieve a Gas Giant composition with radius > 7.5 $R_\oplus$.

Planets with $T_{>630}$ occupying the *sSMRd* likely either (1) experience different core accretion formation conditions and history than $T_{<600}$ planets and therefore never reach a Gas Giant H-He mass fraction, or (2) experience significant post-formation evolutionary effects that strip a significant fraction of accreted H-He gas such that a Gas Giant descends into the mass-radius space of the *sSMRd*.

**Acknowledgements**

This research has made use of the NASA Exoplanet Archive, which is operated by the California Institute of Technology, under contract with the National Aeronautics and Space Administration under the Exoplanet Exploration Program available at: https://exoplanetarchive.ipac.caltech.edu/. This research has made use of NASA's Astrophysics Data System Bibliographic Services available at: https://ui.adsabs.harvard.edu/ and the arXiv eprint service operated by Cornell University available at: https://arxiv.org/ I would like to thank R. Helled for helpful suggestions on a preliminary draft of this paper. I would like to thank the anonymous reviewer for valuable suggestions regarding selection effects in the data.**Data Availability**

The mass, radius, and equilibrium temperatures for the sample of planets discussed in this note was acquired from the NASA Exoplanet Archive via https://exoplanetarchive.ipac.caltech.edu/. The list of planets plotted in Figures 1 and 2 can be provided by the author upon request and are included as supplemental materials following the references.

Szabó G.M., Kálmán S., Borsato L., Hegedűs V., Mészáros S., Szabó R. (2023) Sub-Jovian desert of exoplanets at its boundaries. Parameter dependence along the main sequence. Astronomy and Astrophysics 671, A132. doi:10.1051/0004-6361/202244846

Thorngren D.P., Lee E.J., Lopez E.D. (2023) Removal of Hot Saturns in Mass-Radius Plane by Runaway Mass Loss. The Astrophysical Journal 945, L36. doi:10.3847/2041-8213/acbd35

Trifonov T., Brahm R., Jordán A., Hartogh C., Henning T., Hobson M.J., Schlecker M. et al. (2023) TOI-2525 b and c: A Pair of Massive Warm Giant Planets with Strong Transit Timing Variations Revealed by TESS. The Astronomical Journal 165, 179. doi:10.3847/1538-3881/acba9b

Ulmer-Moll S., Osborn H.P., Tuson A., Egger J.A., Lendl M., Maxted P., Bekkelien A., et al. (2023) TOI-5678b: A 48-day transiting Neptune-mass planet characterized with CHEOPS and HARPS★. Astronomy and Astrophysics 674, A43. doi:10.1051/0004-6361/202245478

Winn J.N. (2010) Transits and Occultations. arXiv e-prints, arXiv:1001.2010. doi:10.48550/arXiv.1001.2010

Youdin A.N. (2011) The Exoplanet Census: A General Method Applied to Kepler. The Astrophysical Journal 742, 38. doi:10.1088/0004-637X/742/1/38
12

**Supplemental Table S1: Teq >630 Planets in the sub-Saturn mass-radius desert    N = 29**

| Planet | Period | M⊕ | R⊕ | Flux/Teq | Reference |
|---|---|---|---|---|---|
| TOI-4010 c | 5.41 | 20.31 | 5.93 | 112 / 907 | Kunimoto et al. 2023 |
| Kepler 56 b | 10.50 | 22.1 | 6.51 | 819 / - | Huber et al. 2013 |
| GJ 436 b | 2.64 | 23.1 | 4.191 | - / 686 | Turner et al. 2016 |
| HAT-P-11 b | 4.89 | 23.4 | 4.36 | 101 / - | Yee et al. 2018 |
| TOI-674 b | 1.98 | 23.6 | 5.25 | 38.4 / 635 | Murgas et al. 2021 |
| HATS-38 b | 4.38 | 24 | 6.88 | 465 / 1294 | Jordán et al. 2020 |
| TOI-1272 b | 3.32 | 24.6 | 4.14 | - / 961 | MacDougal et al. 2022 |
| TOI-1694 b | 3.77 | 26.1 | 5.44 | - / >867 | Van Zandt et al. 2023 |
| Kepler 411 c | 7.83 | 26.4 | 4.421 | - / 838 | Sun et al. 2019 |
| TOI-1728 b | 3.49 | 26.78 | 5.05 | 58 / 767 | Kanodia et al. 2020 |
| TOI-1710 b | 24.28 | 28.3 | 5.34 | - / 687 | König et al. 2022 |
| NGTS-14 A b | 3.54 | 29.2 | 4.98 | - / 1138 | Smith et al. 2021 |
| LTT 9779 b | 0.79 | 29.32 | 4.72 | - / 1978 | Jenkins et al. 2020 |
| LP 714-47 b | 4.05 | 30.8 | 4.7 | - / 700 | Dreizler et al. 2020 |
| K2-27 b | 6.77 | 30.9 | 4.48 | 116 / 902 | Petigura et al. 2017 |
| WASP-166 b | 5.44 | 32.1 | 7.1 | 440 / 1270 | Hellier et al. 2019 |
| K2-19 b | 7.92 | 32.4 | 7.0 | - / >1000 | Petigura et al. 2020 |
| TOI-2498 b | 3.74 | 34.62 | 6.06 | - / 1443 | Frame et al. 2023 |
| HD 89345 b | 11.81 | 35.0 | 7.40 | - / 1089 | Yu et al. 2018 |
| K2-280 b | 19.90 | 37.1 | 7.50 | - / 787 | Nowak et al. 2020 |
| HATS-7 b | 3.19 | 38.1 | 6.31 | - / 1084 | Bakos et al. 2015 |
| TOI-4010 d | 14.71 | 38.15 | 6.18 | 30 / 650 | Kunimoto et al. 2023 |
| WASP-156 b | 3.84 | 40.7 | 5.7 | 150 / 970 | Demangeon et al. 2018 |
| TOI-1288 b | 2.70 | 42 | 5.24 | - / 1266 | Knudstrup et al. 2022 |
| TOI-257 b | 18.39 | 43.9 | 7.16 | 186 / 1027 | Addison et al. 2021 |
| TOI-181 b | 4.53 | 46.17 | 6.96 | 102 / 895 | Mistry et al. 2023 |
| Kepler 101 b | 3.49 | 51.1 | 5.99 | 1240 / - | Bonomo et al. 2014 |
| TOI-532 b | 2.33 | 61.5 | 5.82 | 94 / 867 | Kanodia et al. 2021 |
| CoRoT-8 b | 6.21 | 69.3 | 6.94 | - / 870 | Raetz et al. 2019 |

**Supplemental Table S2: Teq > 630 K,  M⊕ < 20,  R⊕ 4.0 – 7.5 "gas-rich Neptunes"  N = 12**

| Planet | Period | M⊕ | R⊕ | Flux/Teq | Reference |
|---|---|---|---|---|---|
| Kepler 11 e | 32.0 | 6.7 | 4.0 | 32.4 / - | Hadden & Lithwick 2017 |
| K2-79 b | 10.995 | 9.2 | 4.09 | 181 / 1022 | Bonomo et al. 2023 |
| Kepler 79 d | 52.09 | 11.3 | 6.912 | 33.9 / - | Yoffee et al. 2021 |
| K2-19 c | 11.90 | 10.8 | 4.1 | - / 745 | Petigura et al. 2020 |
| K2-32 b | 8.99 | 15.0 | 5.30 | - / 837 | Lillo-Box et al. 2020 |
| HD 106315 c | 21.06 | 15.2 | 4.35 | - / 886 | Barros et al. 2017 |
| Kepler 25 c | 12.72 | 15.2 | 5.22 | 210 / - | Mills et al. 2019 |
| Kepler 18 d | 14.86 | 15.7 | 6.63 | 52 / - | Hadden et al. 2014 |
| TOI-421 c | 16.07 | 16.42 | 5.09 | 34 / 674 | Carleo et al. 2020 |
| HD 219666 b | 6.04 | 16.6 | 4.71 | - / 1073 | Esposito et al. 2019 |
| Kepler 18 c | 7.64 | 18.4 | 5.22 | 126 / - | Hadden et al. 2014 |
| K2-24 b | 20.89 | 19.0 | 5.4 | >60 / > 606 | Petigura et al. 2018 |

**Supplemental Table S3: Gas Giants with Teq>630, M⊕ 20 – 130, R⊕ > 10    N = 40**

| Planet | Period | M⊕ | R⊕ | Flux/Teq | Reference |
|---|---|---|---|---|---|
| WASP-193 B | 6.25 | 44.2 | 16.41 | 413 / 1254 | Barkaoui et al. 2023 |
| WASP-127 b | 4.18 | 52.35 | 14.69 | - / 1400 | Seidel et al. 2020 |
| KELT-11 b | 4.74 | 54.3 | 15.1 | 1453 / 1712 | Beatty et al. 2017 |
| TOI-3976 A b | 6.61 | 55.6 | 12.27 | 469 / 1295 | Yee et al. 2023 |
| HAT-P-18 b | 5.51 | 62.61 | 11.15 | - / 852 | Hartman et al. 2011 |
| TOI-2567 b | 5.98 | 63.9 | 10.93 | 558 / 1352 | Yee et al. 2022 |
| NGTS-12 b | 7.53 | 66.1 | 11.75 | 416 / 1257 | Bryant et al. 2020 |
| HAT-P-12 b | 3.21 | 67.06 | 10.75 | - / 963 | Hartman et al. 2009 |
| TOI-1842 b | 9.57 | 68.0 | 11.7 | - / >1280 | Wittenmyer et al. 2022 |
| TOI-2587 A b | 5.46 | 69.3 | 12.07 | 728 / 1445 | Yee et al. 2023 |
| NGTS-5 b | 3.36 | 72.8 | 12.73 | - / 952 | Eigmüller et al. 2019 |
| HATS-5 b | 4.76 | 75.32 | 10.22 | - / 1025 | Zhou et al. 2014 |
| TOI-2583 A b | 4.52 | 79.5 | 14.46 | 751 / 1456 | Yee et al. 2023 |
| HATS-43 b | 4.39 | 83.0 | 13.23 | - / 1003 | Brahm et al. 2018 |
| TOI-3757 b | 3.44 | 85.3 | 12.0 | 55 / 759 | Kanodia et al. 2022 |
| WASP-131 b | 5.32 | 86 | 13.7 | - / 1460 | Hellier et al. 2017 |
| WASP-160 B b | 3.77 | 88.4 | 12.22 | - / 1119 | Lendl et al. 2019 |
| WASP-39 b | 4.06 | 88.99 | 14.24 | - / 1116 | Faedi et al. 2011 |
| WASP-69 b | 3.87 | 92 | 12.4 | - / 963 | Stassun et al. 2017 |
| HAT-P-19 b | 4.01 | 92.80 | 12.69 | - / 1010 | Hartman et al. 2011 |
| TOI-1296 b | 3.94 | 94.7 | 13.8 | - / 1562 | Moutou 2021 |
| WASP-181 b | 4.52 | 95.0 | 13.27 | - / 1186 | Turner et al. 2019 |
| WASP-117 b | 10.02 | 95 | 11.9 | - / 1001 | Stassun et al. 2017 |
| WASP-83 b | 4.97 | 95 | 11.7 | - / 1120 | Hartman et al. 2015 |
| WASP-21 b | 4.32 | 95.35 | 11.99 | - / 1340 | Bouchy et al. 2010 |
| HAT-51 b | 4.72 | 98.2 | 14.49 | - / 1192 | Hartman et al. 2015 |
| WASP-20 b | 4.90 | 98.84 | 16.39 | - / 1379 | Anderson et al. 2015 |
| WASP-151 b | 4.53 | 100 | 13.31 | - / 1312 | Mocnik et al. 2020 |
| HATS-6 b | 3.33 | 101 | 11.19 | - / 713 | Hartman et al. 2015 |
| K2-295 b | 4.02 | 106 | 10.1 | - / 852 | Smith et al. 2019 |
| TOI-2421 b | 4.35 | 106 | 10.4 | 924 / 1534 | Yee et al. 2022 |
| Kepler 426 b | 3.22 | 108.06 | 12.22 | - / 1300 | Hebrard et al. 2014 |
| HATS-47 b | 3.92 | 117 | 12.52 | 88 / 853 | Hartman et al. 2020 |
| Qatar-8 b | 3.71 | 118 | 14.40 | - / 1457 | Alsubai et al. 2019 |
| HAT-P-58 b | 4.01 | 118 | 14.93 | 1146 / 1622 | Bakos et al. 2021 |
| TOI-2842 b | 3.55 | 118 | 12.85 | 782 / 1471 | Yee et al. 2023 |
| HATS-50 b | 3.83 | 124 | 12.67 |  | Henning et al. 2018 |
| WASP-153 b | 3.33 | 124 | 17.4 | 1400 / 1228 | Nielsen et al. 2019 |
| K2-232 b | 11.17 | 126 | 11.21 | - / 1030 | Brahm et al. 2018 |
| TOI-3912 b | 3.49 | 129 | 14.28 | 872 / 1512 | Yee et al. 2023 |

**Supplemental Table S4: Gas Giants with Teq > 630, M⊕ 20 – 130, R⊕ 7.9 – 10    N = 21**

| Planet | Period | M⊕ | R⊕ | Flux/Teq | Reference |
|---|---|---|---|---|---|
| WASP-139 b | 5.92 | 37.2 | 9.0 | - / 910 | Hellier et al. 2017 |
| HATS-72 b | 7.33 | 39.86 | 8.10 | 50 / 739 | Hartman et al. 2020 |
| Kepler 9 b | 19.24 | 43.4 | 8.29 | 48.2 / - | Borsato et al. 2019 |
| WASP-182 b | 3.38 | 47.0 | 9.53 | - / 1479 | Nielsen et al. 2019 |
| TOI-954 b | 3.68 | 55.3 | 9.55 | 1393 / 1526 | Sha et al. 2021 |
| K2-261 b | 11.63 | 59.8 | 9.53 | 180 / 1046 | Ikwut-ukwa et al. 2020 |
| HD 221416 b | 14.28 | 60.5 | 9.17 | 343 / - | Huber et al. 2019 |
| TOI-2364 b | 4.02 | 71.5 | 8.61 | 237 / 1091 | Yee et al. 2023 |
| HATS-48 A b | 3.13 | 77.2 | 8.97 | 138 / 955 | Hartman et al. 2020 |
| HD 332231 b | 18.71 | 77.6 | 9.72 | 98 / 876 | Dalba et al. 2020 |
| WASP-29 b | 3.92 | 77.9 | 8.88 | - / 970 | Bonomo et al. 2017 |
| TOI-2000 c | 9.13 | 81.7 | 8.14 | 149 / 1038 | Sha et al. 2023 |
| K2-329 b | 12.46 | 82.6 | 8.68 | 46 / 650 | Sha et al. 2021 |
| TOI-3629 b | 3.93 | 83 | 8.3 | 39 / 690 | Cañas et al. 2020 |
| WASP-148 b | 8.80 | 91.2 | 8.47 |  | Almenara et al. 2022 |
| TOI-622 b | 6.40 | 96.3 | 9.24 | 602 / 1388 | Psaridi et al. 2023 |
| TOI-1268 b | 8.16 | 96.40 | 9.1 | - / 919 | Subjak et al. 2022 |
| K2-287 b | 14.89 | 100 | 9.49 | - / 804 | Jordán et al. 2019 |
| HATS-49 b | 4.15 | 112 | 8.57 | 80 / 835 | Hartman et al. 2020 |
| TOI-1298 b | 4.54 | 113 | 9.43 | - / 1388 | Moutou et al. 2021 |
| HD 149026 b | 2.88 | 121 | 8.3 |  | Stassun et al. 2017 |

**Supplemental Table S5: Teq>630, M⊕ 20 -130, R⊕ < 4.0    N = 8**

| Planet | Period | M⊕ | Rearth | Flux/Teq | Reference |
|---|---|---|---|---|---|
| K2-182 b | 4.74 | 20 | 2.69 | 147 / 969 | Akana Murphy et al. 2021 |
| HIP 97166 b | 10.29 | 20.0 | 2.74 | - / 757 | MacDougal et al. 2021 |
| NGTS-4 b | 1.34 | 20.6 | 3.18 | - / 1650 | West et al. 2019 |
| TOI-132 b | 2.11 | 22.40 | 3.42 | 860 / 1395 | Diáz et al. 2020 |
| K2-292 b | 16.98 | 24.5 | 2.63 | 67 / 795 | Luque et al. 2019 |
| Kepler 411 b | 3.01 | 25.6 | 2.401 | - / 1138 | Sun et al. 2019 |
| TOI-2196 b | 1.19 | 26.0 | 3.51 | 2000 / 1860 | Persson et al. 2022 |
| TOI-849 b | 0.77 | 39.09 | 3.444 | - / >1700 | Armstrong et al. 2020 |

**Updated Tables from Russell (2023) T<600 K Catalog (August 22, 2023 Update)**

Table are same as: Russell D.G. (2023) A Catalog of Exoplanets with Equilibrium Temperature less than 600 K. arXiv e-prints. doi:10.48550/arXiv.2303.06214

**Table 3: Rock Terrestrial and Rock-Ice Terrestrial Planets ($f_{H-He}$ < 0.01% by mass)  N = 16**

| Planet | Period | Mass | Radius | Flux / Temp | Reference |
|---|---|---|---|---|---|
| **TRAPPIST-1 h** | 18.77 | 0.326 +/-0.02 | 0.755 +/-0.14 | 0.144 / - | Agol et al. 2021 |
| **TRAPPIST-1 d** | 4.05 | 0.388 +/-0.012 | 0.788 +.011/-.010 | 1.115 / - | Agol et al. 2021 |
| **TRAPPIST-1 e** | 6.10 | 0.692 +/-0.022 | 0.920 +.013/-.012 | 0.646 / - | Agol et al. 2021 |
| **TRAPPIST-1 f** | 9.21 | 1.039 +/-0.031 | 1.045 +.013/-.012 | 0.373 / - | Agol et al. 2021 |
| **TRAPPIST-1 c** | 2.42 | 1.308 +/-0.056 | 1.097 +.014/-.012 | 2.1214 / - | Agol et al. 2021 |
| **TRAPPIST-1 g** | 12.35 | 1.321 +/-0.038 | 1.129 +.015/-.013 | 0.252 / - | Agol et al. 2021 |
| **TRAPPIST-1 b** | 1.51 | 1.374 +/-0.069 | 1.116 +.014/-.012 | 4.153 / - | Agol et al. 2021 |
| **LTT 1445 A c** | 3.12 | 1.54 +.20/-.19 | 1.147 +.055/-.054 | 10.9 / 508 | Winters et al. 2022 |
| **TOI-270 b** | 3.36 | 1.58 +/-0.26 | 1.206 +/-0.039 | - / 581 | Van Eylen et al. 2021 |
| **GJ 1132 b** | 1.63 | 1.66 +/-0.23 | 1.13 +/-0.056 | - / 529 | Bonfils et al. 2018 |
| **GJ 3929 b** | 2.62 | 1.75 +.44/-.45 | 1.09 +/-0.04 | 17.2 / 568 | Beard et al. 2022 |
| **GJ 357 b** | 3.93 | 1.84 +/- 0.31 | 1.217 +.084/-.083 | 12.6 / 525 | Luque et al. 2019 |
| **L98-59 c** | 3.69 | 2.22 +.26/-.25 | 1.385 +.095/-.075 | 12.8 / 553 | Demangeon et al. 2021 |
| **LHS 1478 b** | 1.95 | 2.33 +/-0.20 | 1.242 +.051/-.049 | - / 595 | Soto et al. 2021 |
| **LTT 1445 A b** | 5.36 | 2.87 +.26/-.25 | 1.305 +.066/-.061 | 5.4 / 424 | Winters et al. 2022 |
| **LHS 1140 b** | 24.74 | 6.38 +.46/-.44 | 1.635 +/-0.046 | - / 379 | Lillo-Box et al. 2020 |

**Table 4: Gas-rich, Secondary Envelope, or Supercritical Hydrosphere Terrestrial Planets    N = 18**

| Planet | Period | Mass | Radius | Flux / Temp | Reference |
|---|---|---|---|---|---|
| **L98-59 d** | 7.45 | 1.94 +/- 0.28 | 1.521 +.119/-.098 | 5.01 / 416 | Demangeon et al. 2021 |
| **Kepler 54 c** | 12.07 | 2.10 +.21/-.20 | 1.688 +.054/-.055 | 7.165* / - | Leleu et al. 2023 |
| **Kepler 138 c** | 13.78 | 2.3 +0.6/-0.5 | 1.51 +/-0.04 | 6.8 / 410 | Piaulet et al. 2023 |
| **TOI-244 b** | 7.40 | 2.68 +/-0.30 | 1.52 +/-0.12 | 7.3 / 458 | Castro-González et al. 2023 |
| **HD 260655 c** | 5.71 | 3.09 +/-0.48 | 1.533 +.051/-.046 | 16.1 / 557 | Luque et al. 2022 |
| **Kepler 54 b** | 8.01 | 3.09 +.30/-.31 | 1.856 +/-0.057 | 12.357*/ - | Leleu et al. 2023 |
| **HD 23472 c** | 29.80 | 3.41 +.88/-.81 | 1.87 +0.12/-0.11 | 7.96 / 467 | Barros et al. 2022 |
| **TOI-776 b** | 8.25 | 4.0 +/-0.9 | 1.85 +/-0.13 | 11.5 / 514 | Luque et al. 2021 |
| **G9-40 b** | 5.75 | 4.00 +/-0.63 | 1.900 +/-0.065 | 6.27 / 441 | Luque et al. 2022 |
| **TOI-270 d** | 11.38 | 4.78 +/-0.43 | 2.133 +/-0.058 | - / 387 | Van Eylen et al. 2021 |
| **TOI-1452 b** | 11.06 | 4.82 +/-1.30 | 1.672 +/-0.071 | 1.8 / 326 | Cadieux et al. 2022 |
| **K2-146 b** | 2.64 | 5.77 +/-0.18 | 2.05 +/-0.06 | 20.7 / 534 | Hamann et al. 2019 |
| **TOI-1468 c** | 15.53 | 6.64 +/-0.68 | 2.06 +/-0.044 | 2.15 / 338 | Chaturvedi et al. 2022 |
| **K2-146 c** | 4.00 | 7.49 +/-0.24 | 2.19 +/-0.07 | - / <534 | Hamann et al. 2019 |
| **HD 23472 b** | 17.67 | 8.32 +.78/-.79 | 2.00 +0.11/-0.10 | 16.0 / 543 | Barros et al. 2022 |
| **Kepler 538 b** | 81.74 | 12.9 +/-2.9 | 2.215 +.040/-.034 | 5.07 / 417 | Bonomo et al. 2023 |
| **K2-263 b** | 50.82 | 14.9 +/-2.1 | 2.41 +/-0.12 | 8.17 / 470 | Bonomo et al. 2023 |
| **GJ143 b** | 35.61 | 22.7+2.7/-1.9 | 2.61 +0.17/-0.16 | - / 422 | Dragomir et al. 2019 |

*Planets with stellar flux from Berger et al. 2018

**Table 5: Rock-Ice Giants (Neptunes - ƒ$_{H-He}$ 1.0 – 49.9 % by mass)   N = 37**

| Planet | Period | Mass | Radius | Flux / Temp | Reference |
|---|---|---|---|---|---|
| **HD 191939 d** | 38.35 | 2.8 +/-0.6 | 2.995 +/-0.07 | 14.3 / 540 | Orell-Miguel et al. 2023 |
| **Kepler 289 d** | 66.06 | 4.0 +/-0.9 | 2.68 +/-0.17 | | Schmitt et al. 2014 |
| **Kepler 26 b** | 12.28 | 4.85 +.44/-.42 | 3.22 +/-0.15 | 10.45* / - | Leleu et al. 2023 |
| **Kepler 51 d** | 130.18 | 5.70 +/-1.12 | 9.46 +/-0.16 | 2.466* / - | Libby-Roberts et al. 2020 |
| **TOI-270 c** | 5.66 | 6.15 +/-0.37 | 2.355 +/-0.064 | - / 488 | Van Eylen et al. 2021 |
| **Kepler 305 d** | 16.74 | 6.20 +1.76/-1.34 | 2.76 +/-0.12 | 25.52* / - | Leleu et al. 2023 |
| **LTT 3780 c** | 12.25 | 6.29 +.63/-.61 | 2.42 +/-0.10 | 2.88 / 397 | Nowak et al. 2020 |
| **Kepler 79 e** | 81.06 | 6.3 +/-1.0 | 3.414 +/-0.129 | 18.78* / - | Yofee et al. 2021 |
| **Kepler 87 c** | 191.23 | 6.4 +/-0.8 | 6.14 +/-0.29 | - / 403 | Ofir et al. 2014 |
| **LP 791-18 c** | 4.99 | 7.1 +/-0.7 | 2.438 +/-0.096 | 2.64 / 324 | Peterson et al. 2023 |
| **Kepler 26 c** | 17.25 | 7.48 +.49/-.48 | 3.11 +/-0.14 | 6.63* / - | Leleu et al. 2023 |
| **TOI-178 f** | 15.23 | 7.72 +1.67/-1.62 | 2.287 +/-0.11 | - / 521 | Leleu et al. 2021 |
| **HD 191939 c** | 28.58 | 8.0 +/-0.1 | 3.195 +/-0.075 | 21.0 / 600 | Orell-Miguel et al. 2023 |
| **GJ 1214 b** | 1.58 | 8.17 +/-0.43 | 2.742 +/-.05 | 21.0 / 596 | Cloutier et al. 2021 |
| **Kepler 49 c** | 10.91 | 8.38 +.92/-.89 | 2.444 +.083/-.082 | 12.34* / - | Leleu et al. 2023 |
| **K2-18 b** | 32.94 | 8.63 +/-1.35 | 2.610 +/-0.087 | 1.005 / 255 | Benneke et al. 2019 |
| **TOI-269 b** | 3.70 | 8.80 +/-1.40 | 2.77 +/-0.12 | 19.0 / 531 | Cointepas et al. 2021 |
| **HD 136352 d** | 107.25 | 8.82 +.93/-.92 | 2.562 +.088/-.079 | 5.74 / 431 | Delraz et al. 2021 |
| **HD 73583 c** | 18.88 | 9.70 +1.80/-1.70 | 2.39 +.10/-.09 | 10.2 / 498 | Barragán et al. 2022 |
| **Kepler 49 b** | 7.20 | 9.77 +.94/-.95 | 2.579 +.087/-.086 | 21.48* / - | Leleu et al. 2023 |
| **HD 3167 c** | 29.85 | 10.67 +.85/-.81 | 2.923 +.098/-.109 | - / 565 | Bourrier et al. 2022 |
| **TOI-1759 b** | 18.85 | 10.8 +/-0.15 | 3.14 +/-0.10 | 6.39 / 443 | Espinoza et al. 2022 |
| **Kepler 10 c** | 45.29 | 11.4 +/-1.3 | 2.355 +/-0.022 | 18.68 / 578 | Bonomo et al. 2023 |
| **Kepler 82 b** | 26.44 | 12.15 +.96/-.87 | 4.07 +.24/-.10 | 19.558* / - | Freudenthal et al. 2019 |
| **GJ 3470 b** | 3.34 | 12.58 +/-1.3 | 3.88 +/-0.32 | - / 615 | Kosiarek et al. 2019 |
| **TOI-561 e** | 77.03 | 12.6 +/-1.4 | 2.55 +/-0.11 | 4.96 / 415 | Lacedelli et al. 2022 |
| **TOI-561 d** | 25.71 | 13.2 +1.0/-0.9 | 2.82 +/-0.07 | 21.4 / 598 | Lacedelli et al. 2022 |
| **Kepler 20 d** | 77.61 | 13.4 +3.7/-3.6 | 2.606 +.053/-.039 | 5.71 / 430 | Bonomo et al. 2023 |
| **Kepler 82 c** | 51.54 | 13.9 +1.3/-1.2 | 5.34 +.33/-.13 | 8.035* / - | Freudenthal et al. 2019 |
| **TOI-1246 e** | 37.92 | 14.8 +/-2.3 | 3.78 +/-0.16 | 11 / 462 | Turtelboom et al. 2022 |
| **TOI-3785 b** | 4.67 | 14.95 +4.1/-3.9 | 5.14 +/-0.16 | 19.1 / 582 | Powers et al. 2023 |
| **TOI-1231 b** | 24.25 | 15.4 +/-3.3 | 3.65 +.16/-.15 | 1.93 / 330 | Burt et al. 2021 |
| **K2-24 c** | 42.34 | 15.4 +1.9/-1.8 | 7.5 +/-0.3 | 24 / 606 | Petigura et al. 2018 |
| **TOI-5678 b** | 47.73 | 20 +/- 4 | 4.91 +/- 0.08 | 11.5 / 513 | Ulmer-Moll et al. 2023 |
| **K2-25 b** | 3.48 | 24.5 +5.7/-5.2 | 3.44 +/-0.12 | 9.91 / 494 | Steffansson et al. 2020 |
| **TOI-3884 b** | 4.54 | 32.59 +7.31/-7.38 | 6.43 +/-0.20 | 6.29 / 441 | Libby-Roberts et al. 2023 |
| **HD 95338 b** | 55.09 | 42.44 +2.22/-2.08 | 3.89 +.19/-.20 | 7.42 / 385 | Díaz et al. 2020 |

*Planets with stellar flux from Berger et al. 2018

**Table 6: Gas Giants (Saturns - $f_{H-He}$ 50.1 – 80.0 % by mass)   N = 16**

| Planet | Period | Mass | Radius | Flux / Temp | Reference |
|---|---|---|---|---|---|
| **Kepler 177 c** | 49.41 | 14.7 +2.7/-2.5 | 8.73 +.36/-.34 | 25.4 / - | Vissapragada et al. 2020 |
| **TOI-216 b** | 17.10 | 17.61 +/-0.64 | 7.84 +0.21/-0.19 | 25.9** / - | McKee & Montet 2022 |
| **Kepler 30 d** | 143.34 | 23.1 +/-2.7 | 8.8 +/-0.5 | 1.962* / - | Sanchis-Ojeda et al. 2012 |
| **TOI-2525 b** | 23.29 | 27 +/-2 | 8.68 +/-0.11 | - / 500 | Trifanov et al. 2023 |
| **Kepler 9 c** | 38.99 | 29.9 +1.1/-1.3 | 8.08 +.54/-.41 | 18.903* / - | Borsata et al. 2019 |
| **Kepler 35 b** | 131.46 | 40.36 +/-6.356 | 8.15 +/-0.157 |  | Welsh et al. 2012 |
| **TOI-3984 A b** | 4.35 | 44.0 +8.7/-8.0 | 7.9 +/-0.24 | - / 563 | Cañas et al. 2023 |
| **Kepler 34 b** | 288.82 | 69.92 +3.5/-3.2 | 8.564 +.135/-.157 |  | Welsh et al. 2012 |
| **KOI-1780.01** | 134.46 | 71.0 +11.2/-9.2 | 8.86 +.25/-.24 | 5.70 / - | Vissapragada et al. 2020 |
| **Kepler 16 b** | 228.78 | 105.83 +/-5.08 | 8.449 +/-0.029 |  | Doyle et al. 2011 |
| **NGTS-11 b** | 35.46 | 109 +29/-23 | 9.16 +.31/-.36 | - / 435 | Gill et al. 2020 |
| **K2-139 b** | 28.83 | 123 +26/-24 | 9.06 +.38/-.37 | - / 565 | Barragán et al. 2018 |
| **TOI-530 b** | 6.39 | 127 +29/-32 | 9.3 +/-0.7 | - / 565 | Gan et al. 2022 |
| **TOI-216 c** | 34.55 | 166.9 +/-6 | 10.09 +0.15/-0.13 | 10.10 / 497 | McKee & Montet 2022 |
| **TOI-2525 c** | 49.25 | 209 +/-10 | 10.1 +/-0.1 | - / <500 | Trifanov et al. 2023 |
| **Kepler 167 e** | 1071.23 | 321 +54/-51 | 10.16 +/-0.42 | - / 134 | Chachan et al. 2022 |

*Planets with stellar flux from Berger et al. 2018
**Stellar flux from Kipping et al. 2019

**Table 7: Gas Giants (Jupiters - $f_{H-He}$ 80.1 – 100.0 % by mass)   N = 16**

| Planet | Period | Mass | Radius | Flux / Temp | Reference |
|---|---|---|---|---|---|
| **Kepler 289 c** | 125.85 | 132 +/-17 | 11.59 +/-0.19 | 4.882* / - | Schmitt et al. 2014 |
| **TOI-3235 b** | 2.59 | 211.3 +/-7.95 | 11.16 +/-0.48 | - / 604 | Hobson et al. 2023 |
| **TOI-1899 b** | 29.09 | 212.9 +/-12.71 | 10.86 +/-0.33 | - / 378 | Lin et al. 2023 |
| **CoRoT-9 b** | 95.27 | 267 +/-16 | 11.95 +.84/-.71 | - / 420 | Bonomo et al. 2017 |
| **Kepler 87 b** | 114.74 | 324.2 +/-8.8 | 13.49 +/-0.55 | - / 478 | Ofir et al. 2014 |
| **TOI-5542 b** | 75.12 | 420 +/-32 | 11.31 +.40/-.39 | 9.60 / 441 | Grieves et al. 2022 |
| **KOI-3680 b** | 141.24 | 613 +60/-67 | 11.1 +.7/-.8 | - / 347 | Hébrard et al. 2019 |
| **Kepler 30 c** | 60.32 | 640 +/-50 | 12.3 +/-0.4 | 6.217* / - | Sanchis-Ojeda et al. 2012 |
| **TOI-4127 b** | 56.40 | 731 +/-35 | 12.29 +.44/-.36 | 22.3 / 605 | Gupta et al. 2023- |
| **TOI-4562 b** | 225.12 | 732 +152/-149 | 12.53 +/-0.15 | - / 318 | Heitzmann et al. 2023 |
| **CoRoT-10 b** | 13.24 | 874.00 +/-50.85 | 10.87 +/-0.70 | - / 600 | Bonomo et al. 2010 |
| **TOI-2180 b** | 260.79 | 875.6 +27.7/-25.7 | 11.32 +.25/-.24 | 2.71 / 348 | Dalba et al. 2022 |
| **TOI-2589 b** | 61.63 | 1112 +/-32 | 12.1 +/-0.3 | - / 592 | Brahm et al. 2023 |
| **HD 80606 b** | 111.44 | 1308 +/-49 | 11.24 +/-0.30 | - / 405 | Southworth et al. 2011 |
| **Kepler 1704 b** | 988.88 | 1319 +/-92 | 11.94 +.48/-.46 | 0.342 / 254 | Dalba et al. 2021 |
| **Kepler 1514 b** | 217.83 | 1678 +/-70 | 12.42 +/-0.26 | 3.23 / 388 | Dalba et al. 2021 |

*Planets with stellar flux from Berger et al. 2018